\documentclass[conference]{IEEEtran}
\IEEEoverridecommandlockouts
\usepackage{cite}
\usepackage{amsmath,amssymb,amsfonts}
\usepackage{algorithmic}
\usepackage{graphicx}
\usepackage{textcomp}
\usepackage{xcolor}
\usepackage[a4paper, total={184mm,239mm}]{geometry}

\usepackage{multirow}
\usepackage{booktabs}
\usepackage{hyperref}
\usepackage{subcaption}
\usepackage[export]{adjustbox}
\usepackage[font=small,labelfont=bf]{caption}
\usepackage[ruled,vlined]{algorithm2e}

\usepackage{multirow}
\usepackage{hhline}
\usepackage{makecell}

\hypersetup{
    colorlinks=true,
    linkcolor=blue,
    filecolor=magenta,      
    urlcolor=cyan,
    citecolor = magenta,
    pdfpagemode=FullScreen,
}

\makeatletter
\newcommand*\titleheader[1]{\gdef\@titleheader{#1}}
\AtBeginDocument{%
  \let\st@red@title\@title
  \def\@title{%
    \bgroup\normalfont\large\centering
    \vskip-3.9em
    \@titleheader\par\egroup
    \vskip0.8em\st@red@title}
}
\makeatother

\def\BibTeX{{\rm B\kern-.05em{\sc i\kern-.025em b}\kern-.08em
    T\kern-.1667em\lower.7ex\hbox{E}\kern-.125emX}}

\titleheader{\normalsize{2025 IEEE International Symposium on Circuits and Systems (ISCAS)}}

\title{
\fontsize{22.5}{30.0}\selectfont{Characterization and Mitigation of ADC Noise by Reference Tuning in RRAM-Based Compute-In-Memory\vspace{-5pt}}
}

\author{
Ying-Hao Wei$^*$, Zishen Wan$^*$\thanks{* These authors contributed equally to this work.}, Brian Crafton, Samuel Spetalnick, Arijit Raychowdhury\thanks{This work was supported by CoCoSys, one of seven centers in JUMP2.0, a Semiconductor Research Corporation (SRC) sponsored by DARPA.} \\
\textit{School of Electrical and Computer Engineering, Georgia Institute of Technology, GA, USA}\\
\normalsize \{ywei341, zwan63, bcrafton3, sspetalnick3\}@gatech.edu, arijit.raychowdhury@ece.gatech.edu
\vspace{-5pt}}

\begin{document}


\maketitle

\begin{abstract}
With the escalating demand for power-efficient neural network architectures, non-volatile compute-in-memory designs have garnered significant attention. However, owing to the nature of analog computation, susceptibility to noise remains a critical concern. This study confronts this challenge by introducing a detailed model that incorporates noise factors arising from both ADCs and RRAM devices. The experimental data is derived from a 40nm foundry RRAM test-chip, wherein different reference voltage configurations are applied, each tailored to its respective module. The mean and standard deviation values of HRS and LRS cells are derived through a randomized vector, forming the foundation for noise simulation within our analytical framework. 
Additionally, the study examines the read-disturb effects, shedding light on the potential for accuracy deterioration in neural networks due to extended exposure to high-voltage stress.  This phenomenon is mitigated through the proposed low-voltage read mode. Leveraging our derived comprehensive fault model from the RRAM test-chip, we evaluate CIM noise impact on both supervised learning (time-independent) and reinforcement learning (time-dependent) tasks, and demonstrate the effectiveness of reference tuning to mitigate noise impacts.

\end{abstract}


\section{Introduction}

With the surge in machine learning and deep neural network applications, the demand for enhanced computational capability and efficiency has escalated. Yet, in settings like edge computing environments, power and efficiency become critical constraints. Non-volatile memory devices, distinguished by their negligible power consumption during data retention, emerge as potential answers to this quandary. These encompass RRAM~\cite{RRAM}, MRAM~\cite{xiao2024adapting}, PCM~\cite{Shimeng_PCM}, FeFET~\cite{shou2023see}, and floating-gate devices~\cite{JHaslerRoadmapNeuromorphic}. Beyond the power savings offered by these memory components, compute-in-memory (CIM) architecture~\cite{CiM,SRAM_CIM,peng2019dnn+,chang202373,lele2023heterogeneous} promises further reductions in power consumption.

While RRAM technology boasts several merits, it is also marred by challenges like limited on/off ratios, inter-cell variations, stuck-at-fault and shifts in resistance~\cite{RRAM_SAF,wan2022rram,crafton2022improving}. Beyond RRAM-induced noise, CIM brings forth systemic inaccuracies, such as ADC non-linearity and variance across ADCs.


Recent efforts have attempted to address concerns regarding on/off ratio, ADC non-linearity, module partitioning, and the remapping of defective cells~\cite{bad_cell_remap,huang2022automated,haensch2023compute}. Yet, these solutions often overlook variations between ADCs, which, if neglected, can culminate in challenges like diminished yields and skewed results. On the software aspect, tactics like noise injection during training to bolster robustness~\cite{NeuroSim,noisy_train,wan2023berry} demand a grasp of the actual noise impact. However, a holistic noise analysis necessitates an intricate simulation. Hence, our goal is to create a streamlined model that encompasses primary noise sources. This model aims to gauge the feasibility of an adjustable reference for CIM structures on a module-wise scale, analyzing its influence on diverse tasks. Additionally, given that we operate the test chip under relatively low-stress conditions for the RRAM cells, we demonstrate minimal resistance shifts in these cells, rendering read disturbances almost inconsequential.


\begin{figure}[tb]
    \begin{center}
        \includegraphics[width=\linewidth]{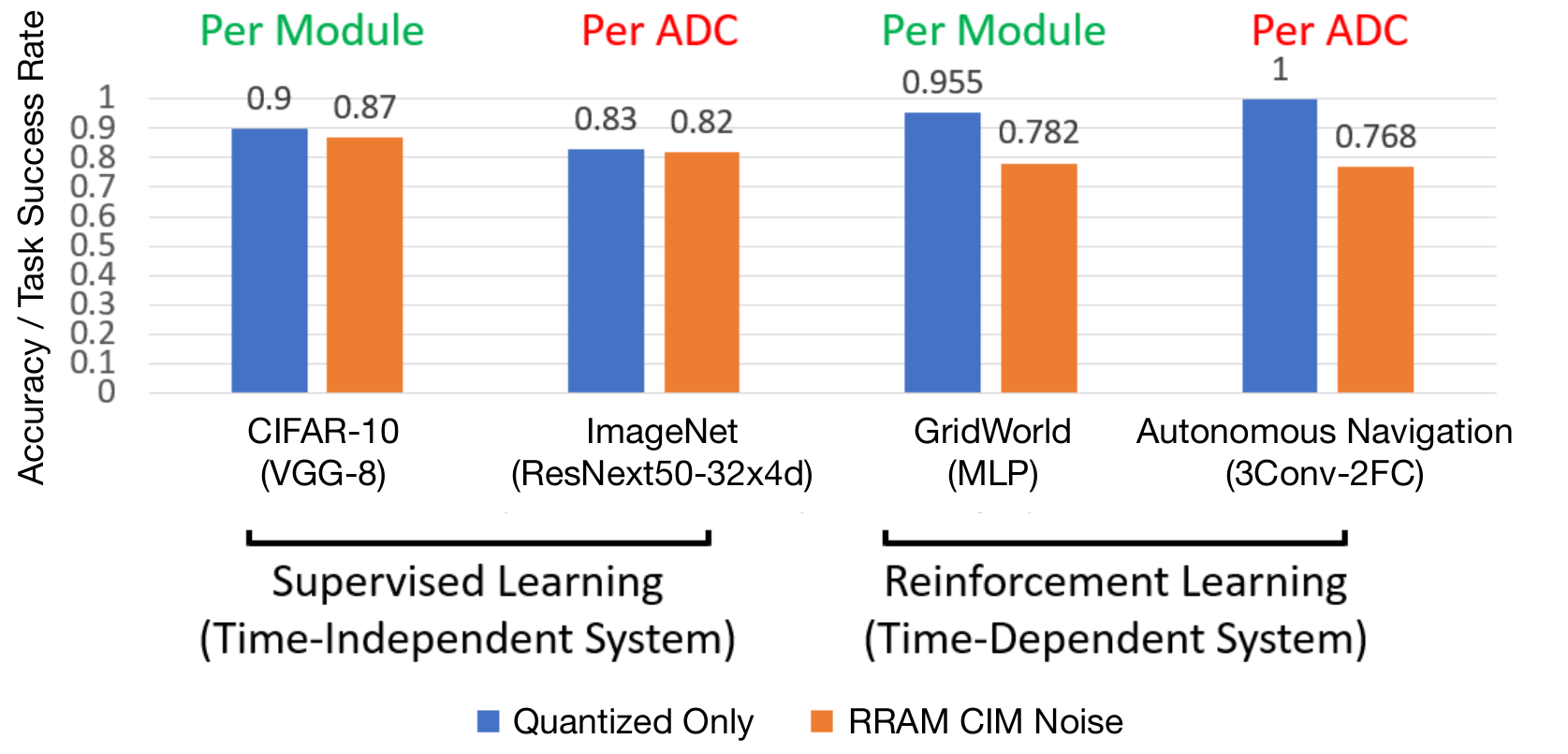}
        \caption{Impact of noises on different scenarios, including supervised learning (time-independent) and unsupervised learning (time-dependent). While some adapted effectively to per-module reference adjustments, others necessitated a more refined per-ADC tuning.}
        \label{fig:intro}
        \vspace{-10pt}
    \end{center}
\end{figure}


    
This paper, therefore, makes the following contributions:
\begin{itemize}
    \item We develop a methodology to encapsulate the behavior of ADCs and RRAM cells, integrating the CIM noise into the cross-layer device-to-application analysis framework.
    \item We analyze and quantify the impact of CIM resistance shifts caused by read disturbances, and evaluate this in the context of supervised (time-independent) and reinforcement (time-dependent) learning scenarios.
    \item We evaluate the accuracy of different workloads and showcase the efficacy of per-module reference tuning.
\end{itemize}

\section{Background and Motivations}
\subsection{Compute-In-Memory}
To enhance the computational efficiency of neural networks, particularly in the context of Vector Matrix Multiplication (VMM), CIM structures are increasingly employed. These structures capitalize on encoding the input vector and distributing it across the wordlines (WLs) of memory cells. Subsequently, when a WL is activated, a greater current flows through the Low Resistance State (LRS) as compared to the High Resistance State (HRS) of the RRAM device. The resulting currents are then aggregated through Kirchhoff's Current Law and quantized through ADCs~\cite{SRAM_CIM,peng2019dnn+}.

\begin{figure}[tb]
    \begin{center}
    \vspace{-5pt}
        \includegraphics[width=.95\linewidth]{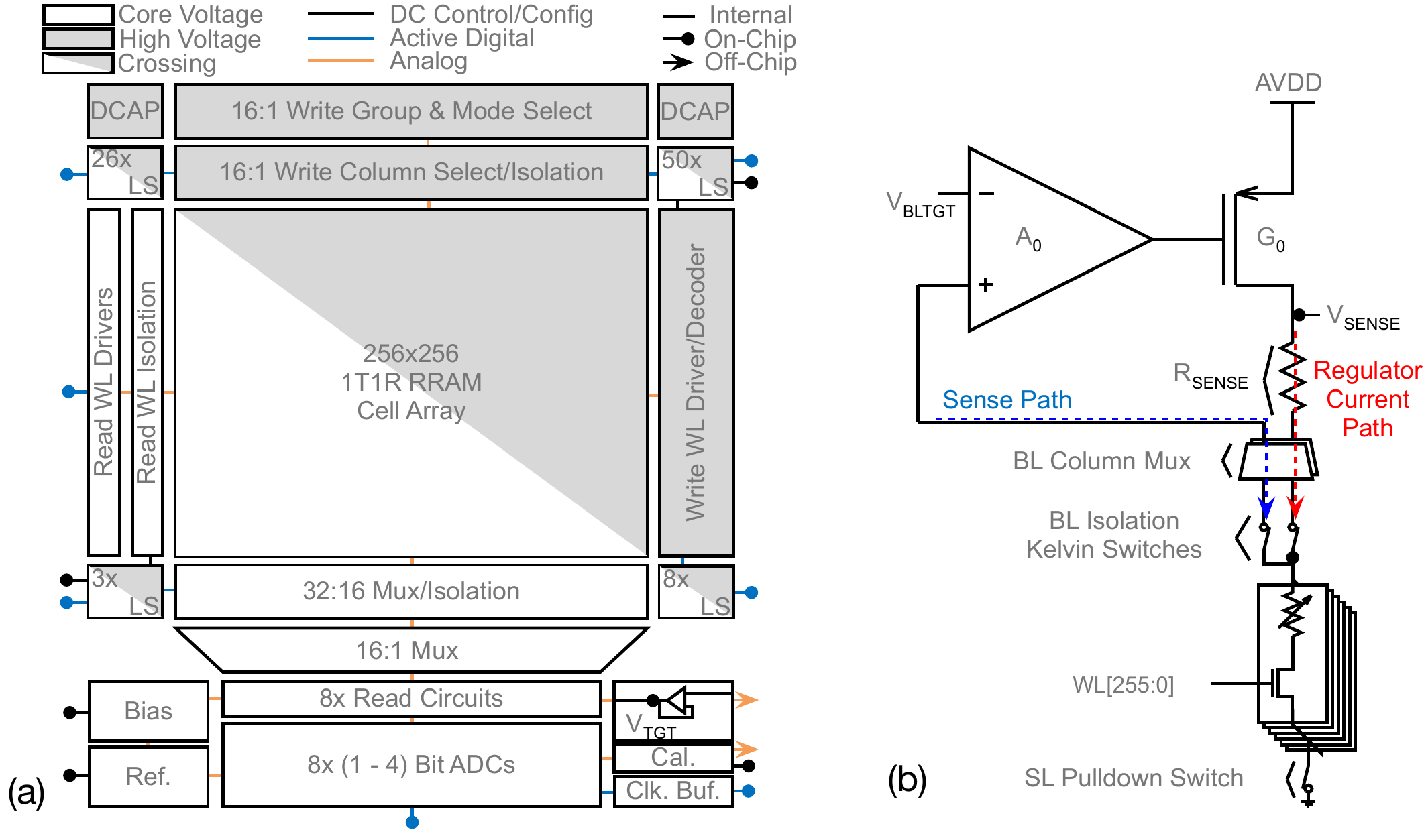}
        \caption{(a) The RRAM macro topology and (b) its symbolic read circuit schematic. Figure adapted from~\cite{RRAM_CiM_original}.}
        \vspace{-20pt}
        \label{fig:read_struct}
    \end{center}
\end{figure}

The experiments conducted in this study adhere to and are validated by the CIM test-chip prototype~\cite{RRAM_CiM_original,RRAM_CiM_original_2,RRAM_resistance_dist,yoon202129}. Fig.~\ref{fig:read_struct} provides an overview of the read circuitry. The output voltage is subsequently decoded using a 4-bit flash ADC, incorporating tunable references. Such a design configuration serves to minimize the required bitline (BL) voltage, thereby reducing both read disturb susceptibility and overall power consumption.


\subsection{Resistive Random Access Memory (RRAM)}
RRAM is a type of non-volatile memory capable of toggling between LRS and HRS through the application of distinct biases, predicated on establishing or interrupting the conduction bridge~\cite{wan2022compute,yu2020compute}. The initial formation of the conduction bridge necessitates a relatively higher bias. This intrinsic characteristic imposes a fundamental constraint on the maximum size of the input vector. Another significant challenge lies in cell-to-cell variance, where resistance inconsistencies further exacerbate the aliasing concern. To quantify the variation and on/off ratio~\cite{RRAM_2}, we will be experimenting on RRAM cells from recent RRAM test-chip prototype on 40nm foundry RRAM arrays~\cite{RRAM_CiM_original,RRAM_CiM_original_2,RRAM_resistance_dist,yoon202129}. 

Moreover, subjecting an RRAM cell to stress induces shifts in its resistance\cite{Shimeng_RRAM_shift}\cite{Shimeng_RRAM_shift_2}. Our read structure, akin to a set operation albeit with differing stress levels, typically exhibits a trend wherein the resistance of HRS shifts towards LRS, while LRS cells tend to remain relatively stable. Such behavior leads to a narrowing of the on-off ratio.

\section{Model For Simulation}
This section describes how we generate the reference for ADCs for optimized accuracy (Sec.~\ref{subsec:adc}) and how the noise is extracted from the test-chip for analysis (Sec.~\ref{subsec:measure}, \ref{subsec:noise}).
\subsection{Generating ADC Configuration}
\label{subsec:adc}
Achieving accuracy in ADC outputs hinges crucially on the precision of the reference settings. We aim to achieve this by introducing random vectors and determining the optimal setting. The process of tuning the internal reference generator encompasses three primary parameters: offset, step size, and BL target voltage, whose designs are also demonstrated in \cite{RRAM_CiM_original}. The offset and step size dictate the reference voltages for the comparators, while the BL target voltage designates the sensing range for different voltage levels.

We define a sequence of nine consecutive cells within a column as an ``acc-9 group." Following this definition, we apply 256 randomly generated 9-bit vectors to every ``acc-9 group" within a matrix of 81 rows and 64 columns, covering the full scope of the 256 columns.

\begin{figure}[tb]
    \begin{center}
    \vspace{-8pt}
        \includegraphics[width=8cm]{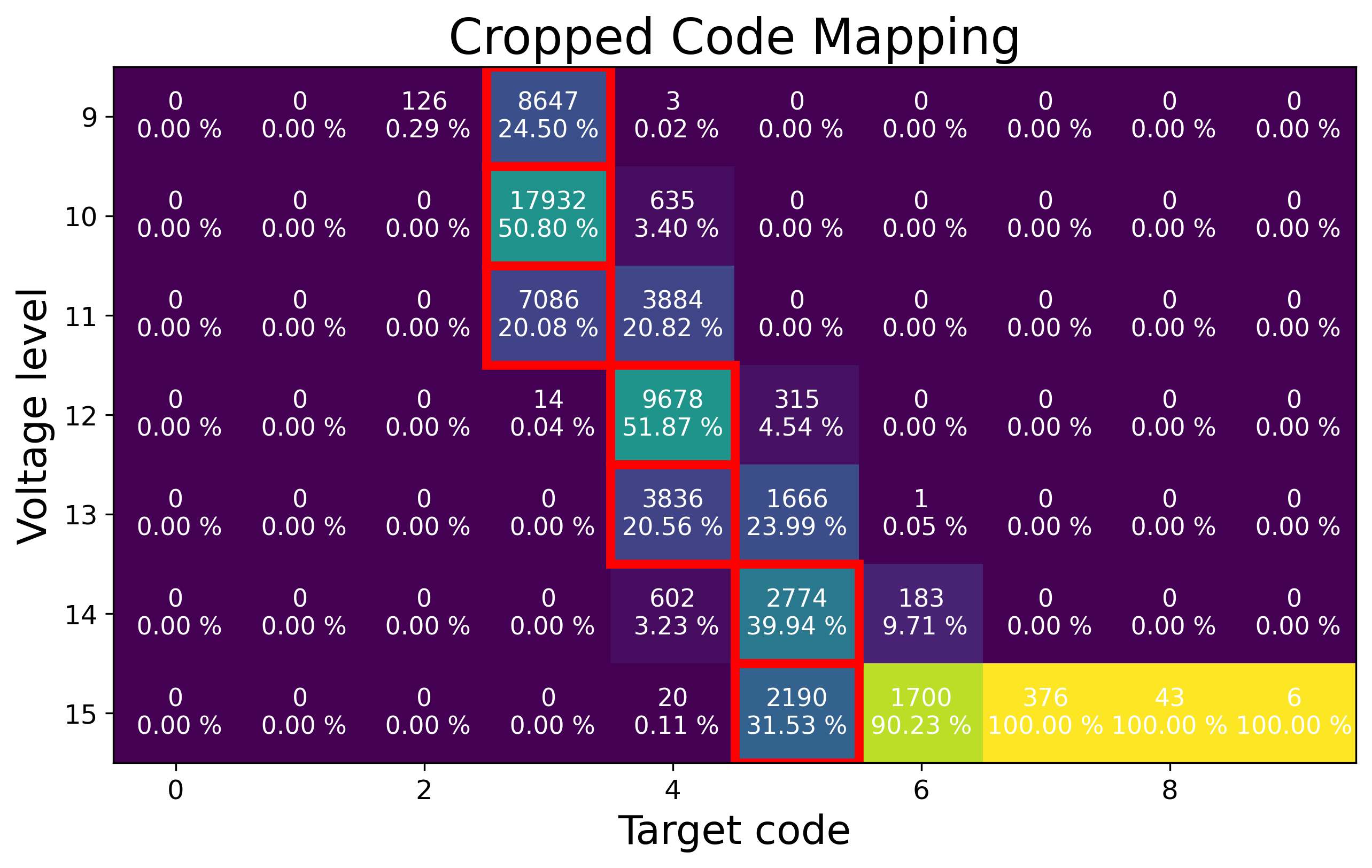}
        \vspace{-2pt}
        \caption{Mapping voltage states to actual value, where x-axis is golden value and y-axis is voltage states. Red boxes indicate that the voltage state is assigned to that specific target code.}
        \vspace{-10pt}
        \label{fig:mod56_assign}
    \end{center}
\end{figure}

Considering the 4-bit ADC's capability to generate 16 distinct output states and the intention to activate a maximum of 9 WLs (leading to 10 unique voltage states), there is a critical need for reliable mapping of ADC outputs to their actual corresponding values. Instead of adopting a fixed mask for this mapping~\cite{RRAM_CiM_original_2}, we utilize the absolute binning approach. For example, in Fig.~\ref{fig:mod56_assign}, the golden value 5 corresponds to voltage level 15 on 2190 times (31.53\% of total golden value 5 occurrence). In contrast, the golden value 6 is represented 1700 times (90.23\% total golden value 6 occurrence). Despite these percentages, voltage level 15 is associated with value 5, as it remains the dominant representation.

In the current setup, the reference is applied globally. This implies that the same step size and offset settings are utilized across all modules. We further probe into the possibility of adopting unique reference settings for varied scales, including individual module and per-ADC setups.

\begin{figure}[tb]
    \begin{center}
    \vspace{-5pt}
        \begin{minipage}{9cm}
            \includegraphics[width=9cm]{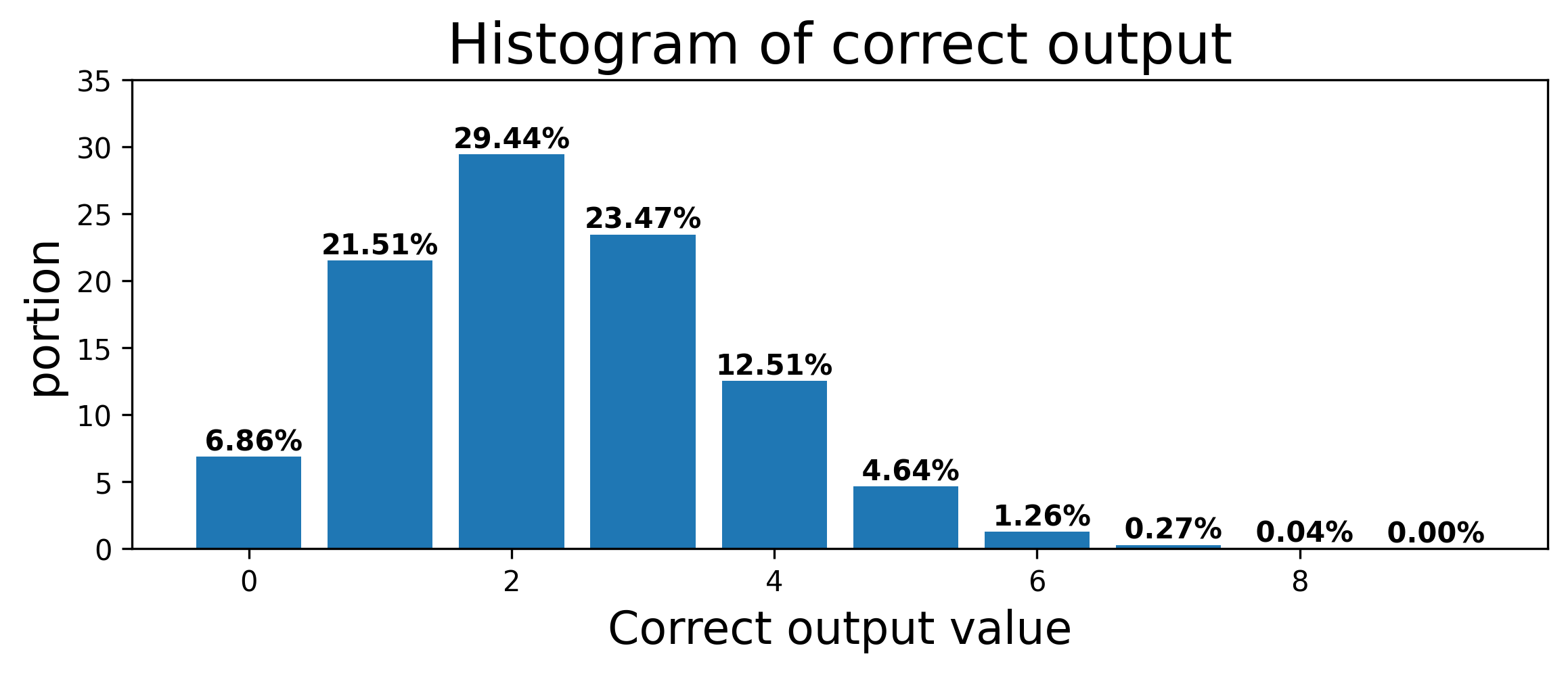}
        \end{minipage}
        \caption{The histogram of the correct output, under a CIM array (9 $\times$ 64 accumulate groups, spams across 8 ADCs) of 50\% HRS and 50\% LRS. The input consists of 256 different combinations among 9 WLs.}
        \label{fig:golden_hist}
        \vspace{-10pt}
    \end{center}
\end{figure}

Fig. \ref{fig:golden_hist} showcases the frequency distribution of each correct response. It's discernible that a significant number of instances are concentrated within the lower sums. This trend becomes particularly pronounced when ReLU layers are incorporated, as they often produce a higher number of zeros. As a result, our efforts primarily concentrate on the precise mapping of the initial 5 and 6 states, while permitting higher sums to attain saturation, as depicted in Fig. \ref{fig:mod56_assign}.

\subsection{Measuring Effective Bit}
\label{subsec:measure}

\begin{figure}[tb]
    \begin{center}
    \vspace{-5pt}
        \includegraphics[width=8.5cm]{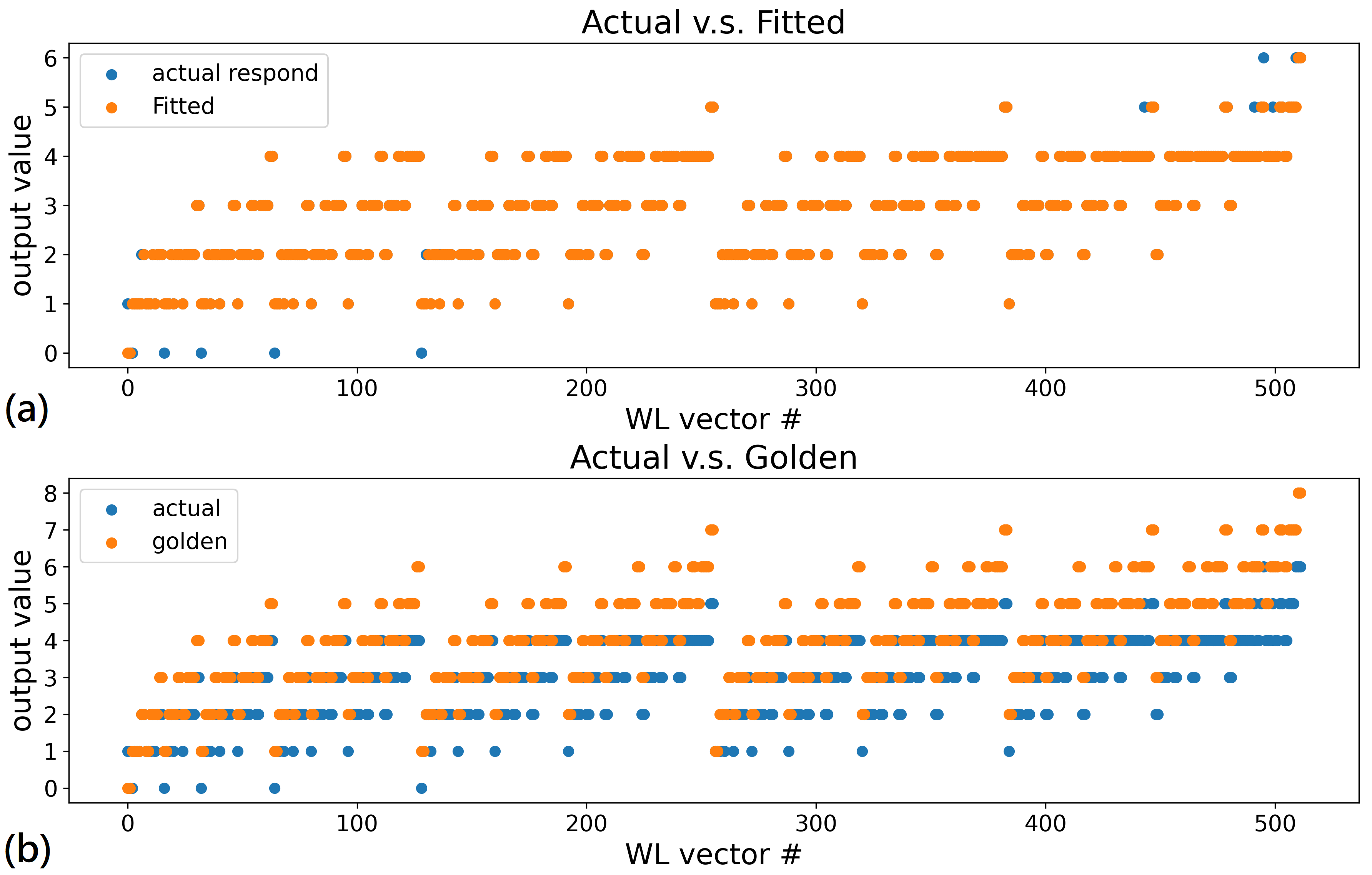}
        \caption{Measuring the effective bits of (a) actual v.s. golden response and (b) actual v.s. fitted response, under 512 WL combinations.}
        \vspace{-10pt}
        \label{fig:AGF_responses}
    \end{center}
\end{figure}
\begin{figure}[tb]
    \begin{center}
        \begin{minipage}{7.5cm}
            \includegraphics[width=7.5cm]{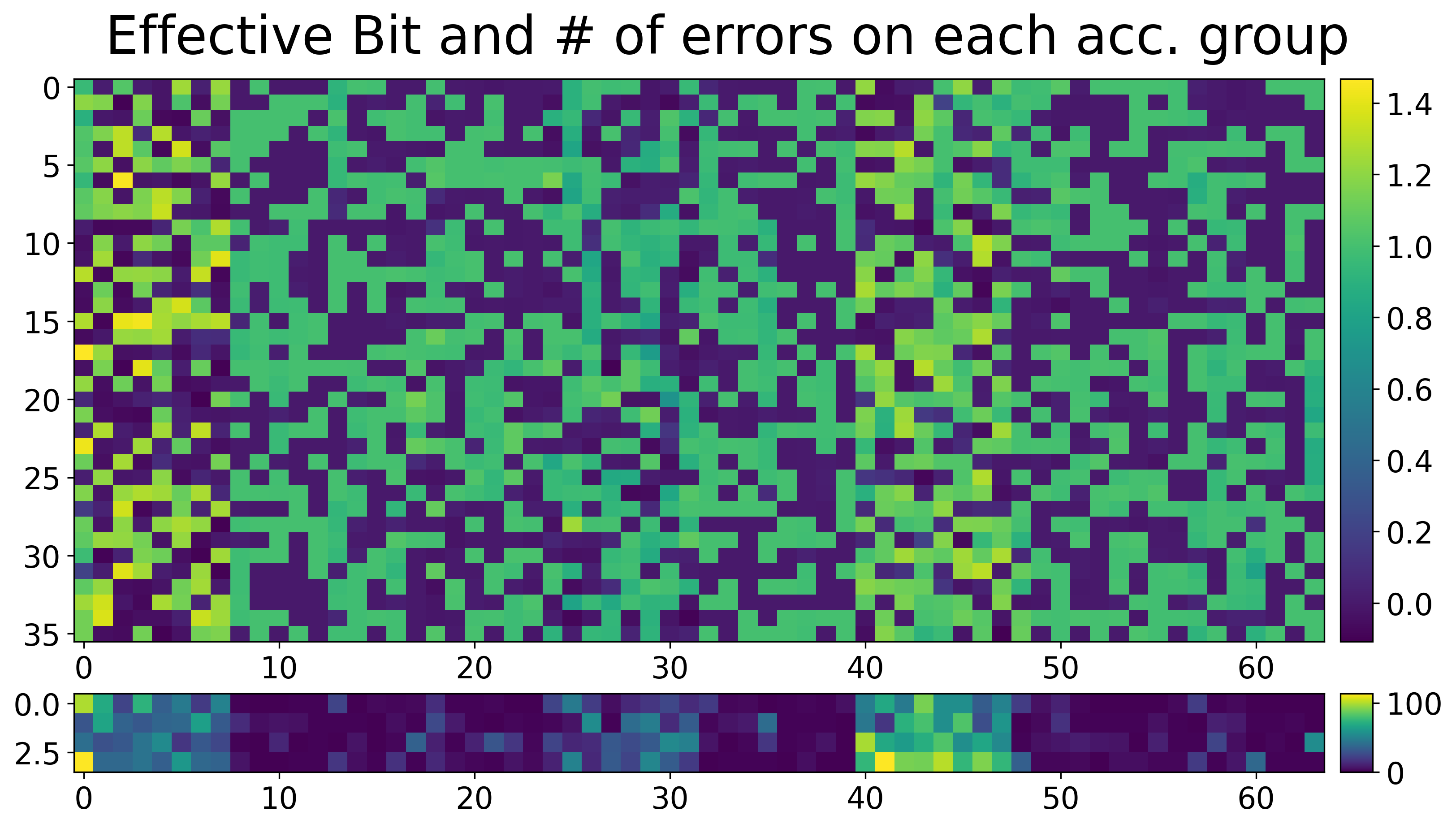}
        \end{minipage}
        \caption{The effective bit (top) and error (bottom) across the module, contains results from all 8 ADCs. The error is calculated by accumulating the difference between golden and fitted output of 256 samples.}
        \label{fig:EBanderr}
        \vspace{-15pt}
    \end{center}
\end{figure}

To construct a realistic model that accurately encompasses both cell variability and ADC effects, we introduce the ``effective bits" concept. This encapsulates the mean and variability of a non-ideal distribution. The derivation of these effective bits comprises multiple stages. Notably, we sampled each accumulation group, comprised of nine cells, using 256 distinct random WL combinations. By analyzing and fitting with minimizing the sum of the absolute difference between input and output patterns, as in Fig.~\ref{fig:AGF_responses}, we pinpoint the effective bit for each cell within an accumulation group. We'll reference this element as the static error component. It's crucial to emphasize that the fitting methodology isn't without its imperfections and introduces an auxiliary error, as depicted in Fig.~\ref{fig:AGF_responses}\textcolor{blue}{(b)}. Hence, we aggregate the distribution of this residual error and assimilate it as a dynamic error source. The distribution of effective bits throughout a module is illustrated in Fig.~\ref{fig:EBanderr}, underscoring the profound influence of disparate ADCs on both the effective bits and their accompanying errors.

\subsection{Noise Injection Methodology}
\label{subsec:noise}
The static noise is initially injected, as follows. Firstly, we retrieve the 0/1 effective bit distribution from 10 distinct modules (encompassing 80 ADCs). We then harness the mean and standard deviation to spawn random normal distributions for both 0s and 1s. It's noteworthy that these distributions are curated independently. Finally, we replace the original bits with the derived effective bits. In some cases, tuning references on a per-module basis (involving 8 ADCs) might not be successful, necessitating a shift to per-ADC tuning. The methodology remains largely analogous, albeit with the generation of configurations tailored for each ADC.

\section{Impact Of CIM Read Disturb}
This section quantifies (Sec.~\ref{subsec:quantify}) and evaluates (Sec.~\ref{subsec:read_disturb}) the potential impact of read disturb under different scenarios. 

\subsection{Quantifying Read Disturb}
\label{subsec:quantify}

\begin{figure}[t!]
    \begin{center}
    \vspace{-5pt}
        \begin{minipage}{8.5cm}
            \includegraphics[width=8cm]{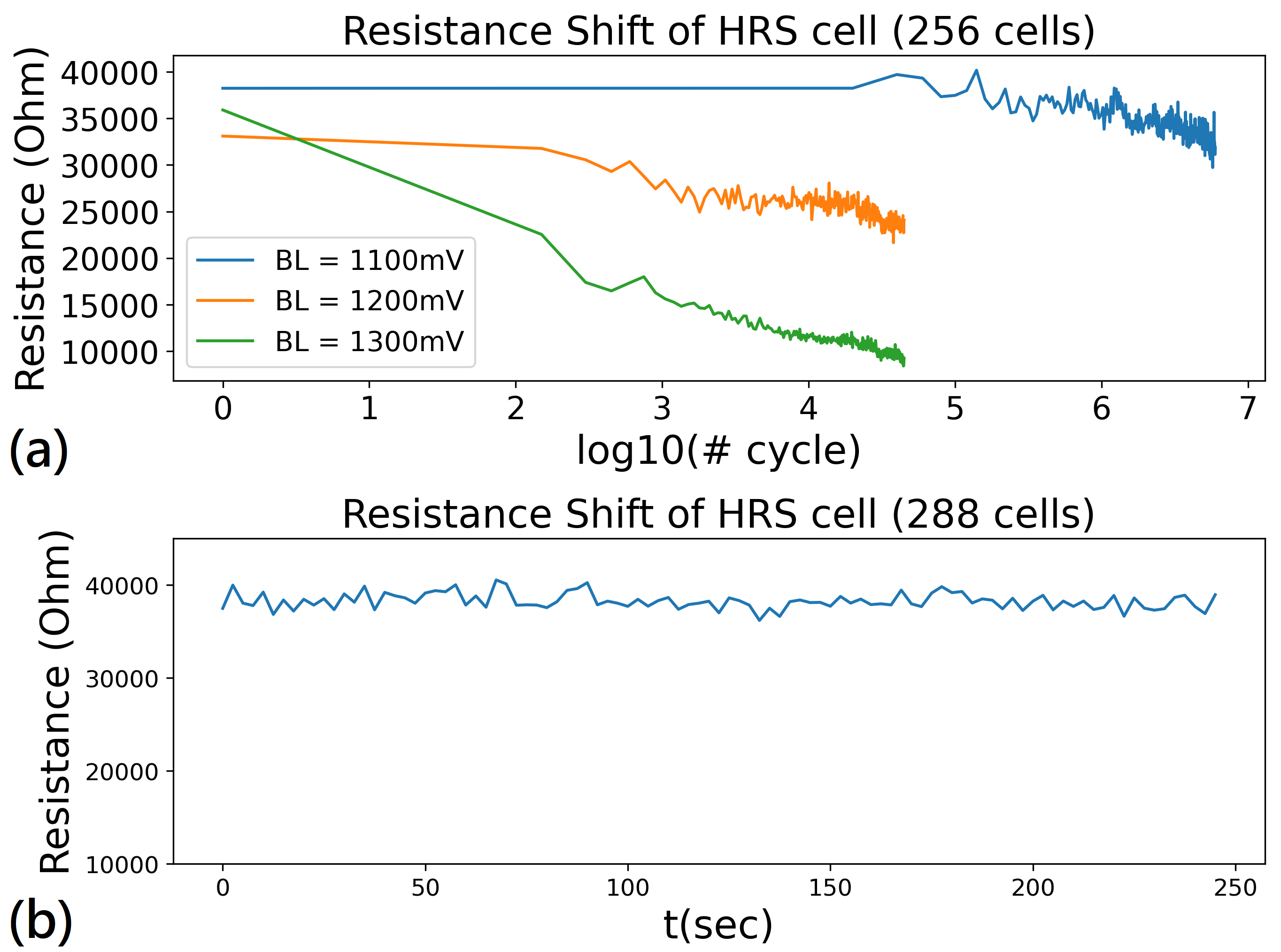}
        \end{minipage}
        \caption{(a) Resistance shift under variance BL voltage and Fixed WL voltage under normal temperature. A cycle is (5/64M) second. (b) Resistance shifts under WL=1.1V, BL=0.3V, over 250 seconds.}
        \label{fig:R_shift_VS_BL}
        \vspace{-18pt}
    \end{center}
\end{figure}

In the CIM architecture, the reading process bears significant resemblance to a set operation. This makes HRS cells particularly prone to read disturbances. We observe that HRS cells tend to undergo shifts, while LRS cells mostly maintain their stability over time. Fig.~\ref{fig:R_shift_VS_BL} shows the effects of different stress levels applied at varied BL voltages, keeping the WL voltage consistently at 1.1V. It's worth noting that the bias voltage used for reading is considerably lower, often around 100mV and peaking at 300mV. The results, as illustrated in Fig.~\ref{fig:R_shift_VS_BL}, confirm a minimal resistance shift for periods exceeding 250 seconds under ambient temperature conditions.


\subsection{Read Disturb Impact of DNN Models}
\label{subsec:read_disturb}
\begin{figure}[t!]
\centering
    \begin{subfigure}[t]{.48\linewidth}
          \includegraphics[width=\linewidth]{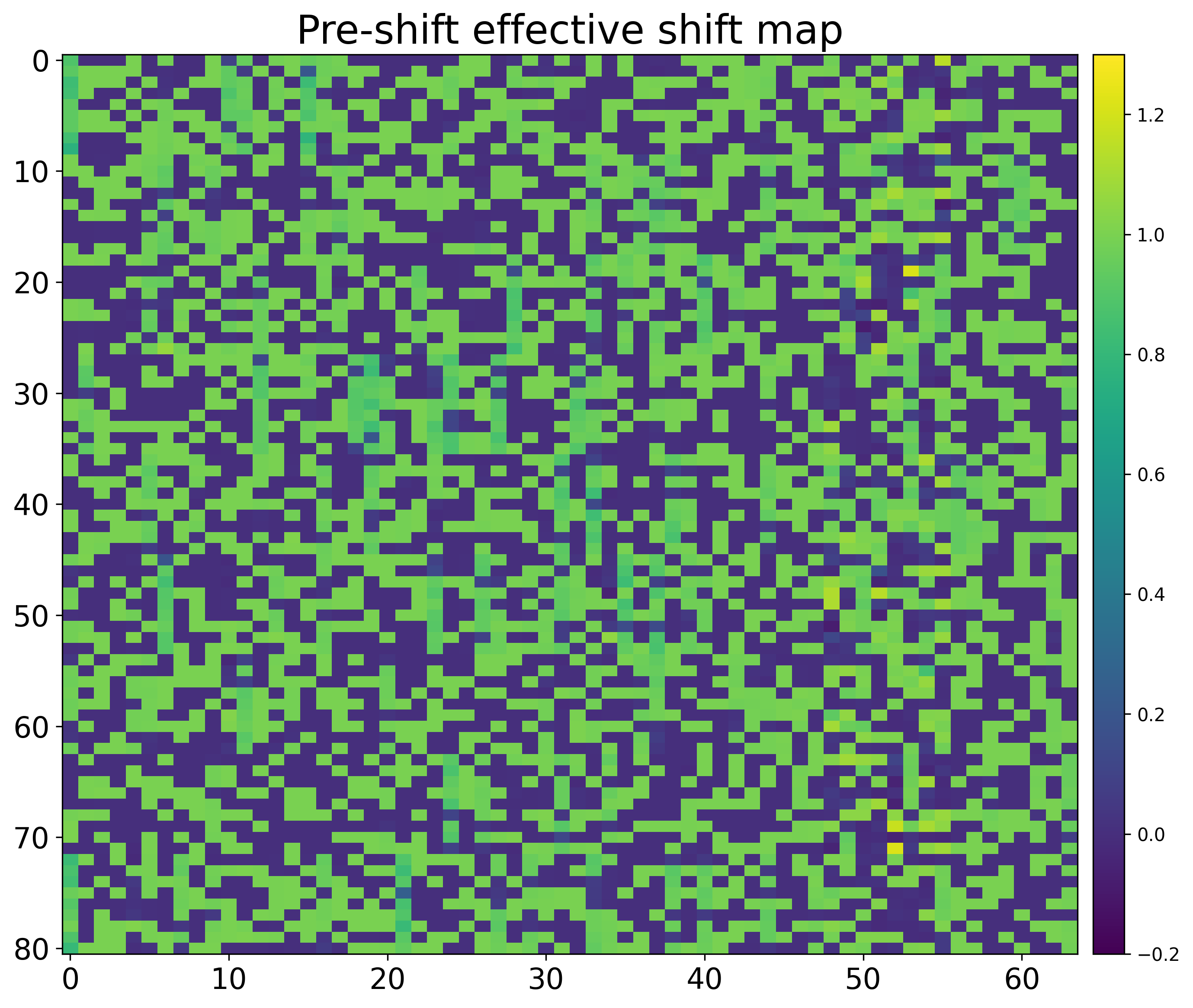}
          \vskip -0.1in
          \caption{}
          \label{fig:eb_preshift}
    \end{subfigure}
    \hskip -0.05in
        \begin{subfigure}[t]{.48\linewidth}
          \includegraphics[width=\linewidth]{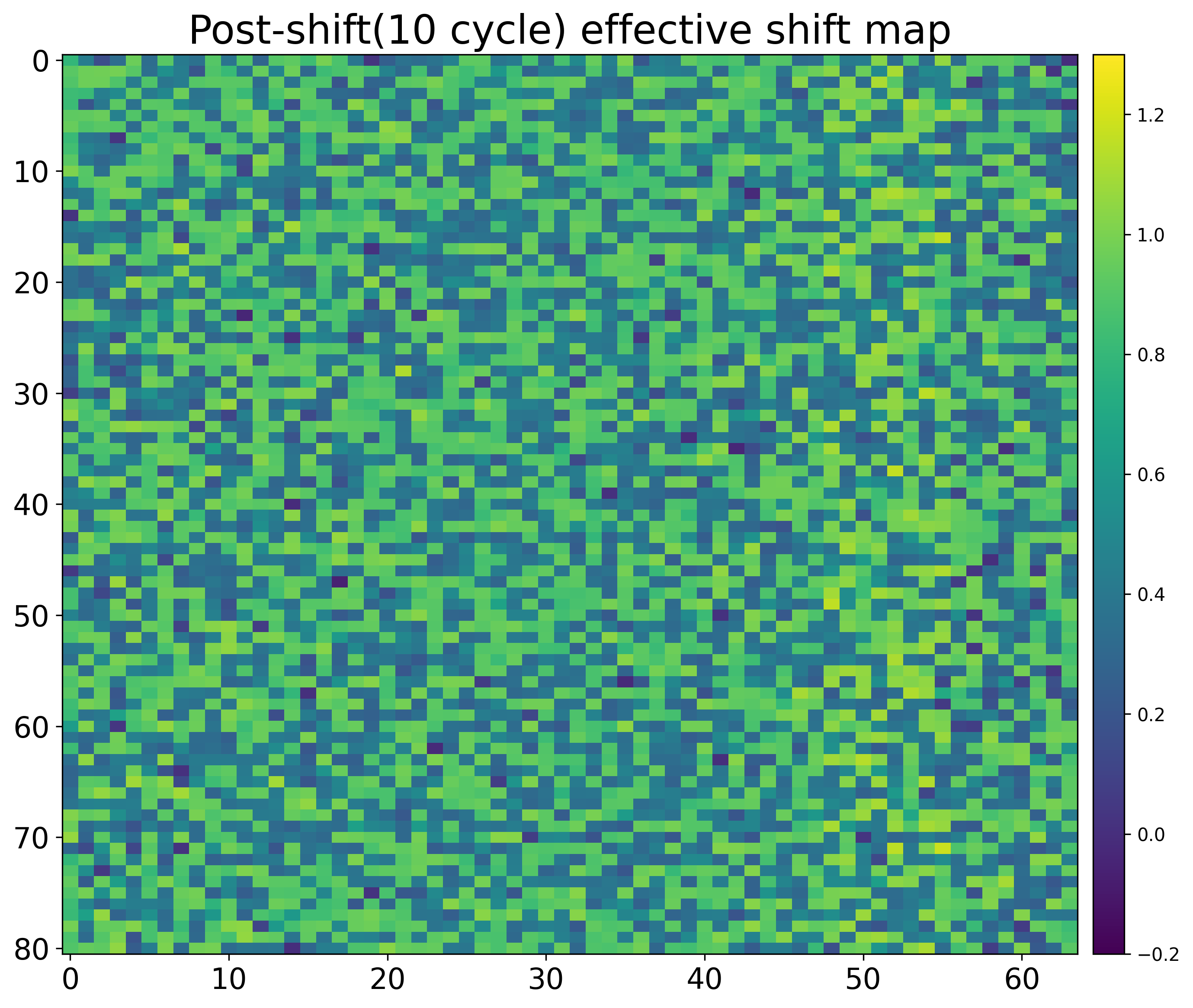}
          \vskip -0.1in
          \caption{}
          \label{fig:eb_postshift}
    \end{subfigure}
    \vspace{-0.03in}
    \caption{(a) Effective bit map for pre-shift, (b) Effective bit map for post-shift.}
    \vskip -0.15in
    \label{fig:eb_map}
\end{figure}

\begin{figure}[t!]
    \begin{center}
    \vspace{-5pt}
        \begin{minipage}{8cm}
            \includegraphics[width=8cm]{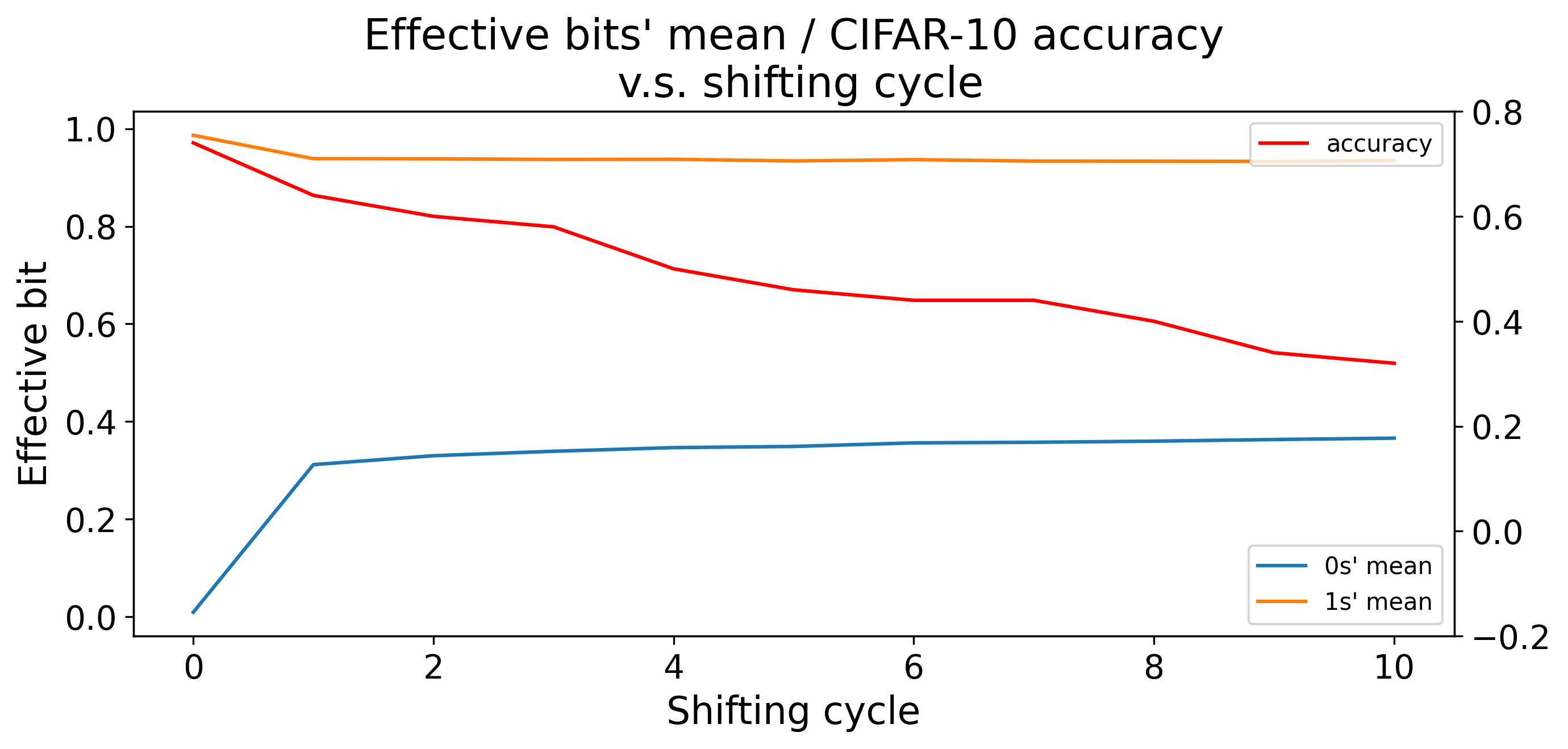}
        \end{minipage}
        \caption{Effective bit mean of HRS(0) (orange)/LRS(1) (blue) and accuracy of CIFAR-10 (red) v.s. cycle.}
        \label{fig:eb_vs_cycle}
        \vspace{-10pt}
    \end{center}
\end{figure}

We specifically opt for a module equipped with a relatively linear ADC setup to assess the impact of shifting. This choice stems from the fact that more precise modules tend to exhibit a more discernible disparity in noise levels when compared to initially noisy modules. To speed up the read disturb process, we select higher stress, and hence, in Fig.~\ref{fig:eb_map}, we present the effective bit values before and after shifting, conducted under conditions involving a 1300mV BL and 10 cycles of 50k stresses, distributed uniformly across all cells. The trend in effective bit means across shifting cycles is portrayed in Fig.~\ref{fig:eb_vs_cycle}, further affirming the greater susceptibility of HRSs to read disturbance. 


Following the determination of effective bit shifts, we incorporate the resistance shifts into VGG-8, based on the parameters measured from module. We observe that the shifts result in CIFAR-10 accuracy drop from 87\% down to 66\%.

\section{Impact of CIM Noise: Device to Application}

This section explores CIM noise impact across various applications, including supervised learning (Sec.~\ref{subsec:supervised}) and reinforcement learning (Sec.~\ref{subsec:RL}) paradigms with varying complexities.

\subsection{Supervised Learning (Time-Independent System)}
\label{subsec:supervised}
Supervised learning involves training a model using labeled data, where the correct answer is provided during training. We train the models of VGG-8 (on CIFAR-10) and ResNeXt-50$\times$32d~\cite{resnext50_32x4d_paper} (on Imagenet-1K), utilizing a modified version of Neurosim~\cite{NeuroSim}. This approach incorporates layer scaling and WAGE quantization~\cite{WAGE_originial_paper}. Tab.~\ref{tb:supervise_learning} presents the accuracy achieved with different bit configurations for weights and activations.


When applied with per-module reference tuning on ResNeXt50-32$\times$4d (on ImageNet), the network encounters failures, suggesting that per-module level reference setup might be inadequate. However, opting for per-ADC level tuning allows us to maintain accuracy. With quantization, the accuracy stood at 83\%, which marginally dipped to 82\% post-noise injection.

\begin{table}[!htb]
    \caption{Experimental configurations and accuracy of CIFAR-10/VGG-8 and Imagenet/ResNeXT50-32$\times$4d models.}
    \begin{subtable}{.5\linewidth}
        \centering
        \begin{tabular}{| c   c | c | c | }
        \hline
  \multicolumn{2}{|c|}{Quantize}&\multicolumn{2}{c|}{Weight} \\  \cline{3-4}
  \multicolumn{2}{|c|}{only}    &    4b     &    5b          \\  \hline
         \parbox[t]{2mm}{\multirow{4}{*}{\rotatebox[origin=c]{90}{Activate}}}
 &\multicolumn{1}{|c|}{5b}& 73.7\%  &  89.0\%         \\  \cline{2-4}
 &\multicolumn{1}{|c|}{6b}& 75.1\%  &  89.6\%         \\  \cline{2-4}
 &\multicolumn{1}{|c|}{7b}& 74.7\%  &  90.0\%         \\  \cline{2-4}
 &\multicolumn{1}{|c|}{8b}& 75.0\%  &  89.9\%         \\  \hhline{|====|}
  \multicolumn{2}{|c}{baseline}    &  \multicolumn{2}{|c|}{90.03\%} \\ \hline
  \multicolumn{2}{|c}{\makecell{noise-injected\\(per-module)}}&\multicolumn{2}{|c|}{87.7\%} \\ \hline
        
        \end{tabular}
        \vspace{5pt}
        \caption{CIFAR-10 / VGG-8}
    \end{subtable}%
        \begin{subtable}{.5\linewidth}
        \centering
        \begin{tabular}{| c   c | c | c | }
        \hline
  \multicolumn{2}{|c|}{Quantize}&\multicolumn{2}{c|}{Weight}  \\  \cline{3-4}
  \multicolumn{2}{|c|}{only}    &    6b     &    8b           \\  \hline
         \parbox[t]{2mm}{\multirow{4}{*}{\rotatebox[origin=c]{90}{Activate}}}
 &\multicolumn{1}{|c|}{6b} &  0\%  &   0\%              \\  \cline{2-4}
 &\multicolumn{1}{|c|}{8b} & 29\%  &  75\%              \\  \cline{2-4}
 &\multicolumn{1}{|c|}{10b}& 35\%  &  78\%              \\  \cline{2-4}
 &\multicolumn{1}{|c|}{12b}& 36\%  &  78\%              \\  \hhline{|====|}
  \multicolumn{2}{|c}{\makecell{baseline\\(reduced-set)}}    &  \multicolumn{2}{|c|}{83\%} \\ \hline
  \multicolumn{2}{|c}{\makecell{noise-injected\\(per-ADC,\\reduced-set)}}
  &\multicolumn{2}{|c|}{82\%} \\ \hline     
        \end{tabular}
        \vspace{5pt}
        \caption{Imagenet-1k / ResNeXT50-32x4d}
    \end{subtable}%
    \label{tb:supervise_learning}
    \vspace{-10pt}
\end{table}

\subsection{Reinforcement Learning (Time-Dependent System)}
\label{subsec:RL}

\begin{table}[t!]
\caption{GridWorld win rate/mean step under different bit settings.}
\centering
    \begin{tabular}{| c   c | c | c | c | }
    \hline
\multicolumn{2}{|c|}{Quantize}&\multicolumn{3}{c|}{Weight}     \\  \cline{3-5}
\multicolumn{2}{|c|}{only}    &  2b    &    4b    &    6b       \\ \hline
\parbox[t]{2mm}{\multirow{3}{*}{\rotatebox[origin=c]{90}{Activate}}}
&\multicolumn{1}{|c|}{5b}& 4.8\% / 3.66 &  4.8\% / 3.66 & 94.9\% / 6.51 \\ \cline{2-5}
&\multicolumn{1}{|c|}{6b}& 4.8\% / 3.66 & 95.5\% / 6.54 & 95.3\% / 6.52 \\ \cline{2-5}
&\multicolumn{1}{|c|}{8b}& 4.8\% / 3.66 & 95.5\% / 6.55 & 95.3\% / 6.52 \\ \hhline{|=====|}
\multicolumn{2}{|c}{\makecell{Noise-injected\\(per-module)}}&\multicolumn{3}{|c|}{78.2\% / 6.17} \\ \hline
    \end{tabular}
    \label{tb:gridworld_WA_VS_acc}
\end{table}

\begin{table}[t!]
\caption{Drone autonomous navigation performance evaluation.}
\centering
    \begin{tabular}{| c | c | }
    \hline
    Baseline MSF      & 1215.65m \\ \hline
    Quantize Only MSF    &  899.38m \\ \hline
    Per-module tuning MSF &  $<$ 20m  \\ \hline
    Per-ADC tuning MSF  & 690.99m  \\ \hline
    \end{tabular}
    \label{tb:PEDRA_performance}
    \vspace{-10pt}
\end{table}

Reinforcement learning (RL) empowers agents to optimize decision-making by interacting with environments and receiving performance feedback as rewards or penalties. Over iterations, agents refine strategies to maximize rewards~\cite{krishnan2019quarl}. We evaluate the impact of CIM noise on two RL-based template problems: Grid World and drone autonomous navigation.

 
 
We begin our experimentation with a simpler but popular navigation problem of Grid World~\cite{wan2021analyzing,wan2022frl}. The environment consists of a $n\times n$ grid, and the goal of agent is to reach the goal and avoid getting trapped in the hole. We trained an MLP network with two fully connected layers, where inputs encompass 8 values that indicate possible collisions and goal directions. Tab.~\ref{tb:gridworld_WA_VS_acc} details the average win rates and steps taken across 10K missions using the quantized network on CIM.

We then experiment on a more complex drone navigation problem in 3D realistic environments~\cite{PEDRA_paper,wan2022analyzing,hsiao2023mavfi}. During the task, the drone is required to navigate across the environment avoiding obstacles. The monocular image captured from the front-facing camera is taken as the state of the RL problem. We use a perception-based probabilistic action space with 25 actions, and 8-bit quantized C3F2 model with 3 Conv layers and 2 FC layers. To quantify the performance, we use Mean Safe Flight (MSF) which is the average distance traveled by the drone before collision. The results are shown in Tab~\ref{tb:PEDRA_performance}, it is well observed that RRAM noise degrades both systems' performance and per-module/per-ADC tuning is effective in maintaining system robustness.




\section{Conclusion}
Noise in CIM remains a critical concern. To address that, we build up a simulation that extracts the noises from the 40nm RRAM test-chip. Through the analysis, we infer several key findings: complex problems are more susceptible to noise and need more precision on the hardware, especially those that are time-dependent, are more susceptible to noise than supervised learning paradigms. The adverse effects of read disturbances can be mitigated when operations occur in low-read scenarios.




\bibliographystyle{ieeetr}
\bibliography{ref}

\end{document}